\documentclass{article}

\usepackage{arxiv}

\usepackage[utf8]{inputenc} 
\usepackage[T1]{fontenc}    
\usepackage{hyperref}       
\usepackage{url}            
\usepackage{booktabs}       
\usepackage{amsfonts}       
\usepackage{nicefrac}       
\usepackage{microtype}      
\usepackage{lipsum}
\usepackage{graphicx}
\usepackage{subcaption}

\usepackage[
    style=numeric-comp,
    bibstyle=numeric,
    sorting=none,
    natbib=true,
    citestyle=numeric,
    backend=biber 
]{biblatex} 
\title{Introducing the Perturbative Solution of the Inter-Channel Stimulated Raman Scattering in Single-Mode Optical Fibers}

\author{
 Andrea D'Amico \\
  Department of Electronics and Telecommunications \\
  Politecnico di Torino\\
  Turin, Italy \\
  \texttt{andrea.damico@polito.it} \\
   \And
 Giacomo Borraccini \\
  Department of Electronics and Telecommunications \\
  Politecnico di Torino\\
  Turin, Italy \\
  \texttt{giacomo.borraccini@polito.it} \\
  \And
 Vittorio Curri \\
  Department of Electronics and Telecommunications \\
  Politecnico di Torino\\
  Turin, Italy \\
  \texttt{vittorio.curri@polito.it} \\
}

\begin{document}
\maketitle
\begin{abstract}
The continuously increasing IP data traffic demand, with geometrical growth rate exceeding 26\%, requires a large transmission capacity increment from the fiber optical infrastructure.
As the deploy of new fiber cables requires extensive investments, the development of multi-band amplifiers and transceivers, already available as prototypes, is progressively considered towards the entire low-loss single-mode bandwidth beyond the 5~THz C-band.
In this perspective, an adequate handling of the variations along the frequency of the fiber physical features becomes crucial for the fiber propagation modeling in multi-band wavelength division multiplexing (WDM) channel comb transmission scenarios. 
In particular, the inter-channel stimulated Raman scattering (SRS) is the fundamental inter-band effect in this context.
The SRS effect on the WDM comb propagated through a single-mode optical fiber is described by a set of ordinary differential equations (ODEs).
To date, an exact solution of the SRS ODEs has not been proposed, and in the literature numerical solutions or approximations have been considered in order to take into account this effect. 

In this work, a perturbative solution of the SRS ODEs is presented enabling an efficient trade-off between the target accuracy and the computational time. 
Considering a C+L+S transmission scenario, the perturbative expansion up to the 2$^{nd}$ order ensures an excellent accuracy.
Whereas, in an U-to-E transmission scenario, the 3$^{rd}$ order is required in order to reach an equivalent accuracy.
\end{abstract}


\section{Introduction}
The demand for IP data traffic is ever increasing and medium term authoritative forecasts envision a geometrical growth exceeding 26\% as compound annual growth rate (CAGR) on the average, with a much larger figure for some network segments \cite{cisco2019}.
To support data transport, Wavelength Division Multiplexed (WDM) fiber optics transmission and networking using dual-polarization coherent optical technologies is expanding from core- and metro-networks to the access, 5G $x$-hauling \cite{5g} and inter- and intra-datacenter connections.
In this perspective, the fiber optical infrastructure must progressively support the continuously increasing data transport. 
In the telecommunication framework, the deployment of new infrastructures requires large CAPEX investments, and in optical networks installing new cables is particularly expensive \cite{8966280}.

The largest portion of installed and under deployment fiber variety is the standard single mode fiber (SSMF) made of purified glass (ITU-T G.652D fiber) \cite{CRU2018}, that is characterized by low loss profile, below 0.4 dB/km, in the single-mode spectral region, in absence of the water absorption peaks.
In particular, the overall available transmission bandwidth of already installed cables exceeds 50 THz: the U, L, C, S, E and O bands.
Consequently, the exploitation of the entire transmission bandwidth represents an interesting solution for the total capacity increasing, maximizing returns from CAPEX investments \cite{ferrari2020assessment,9893169}.

Currently, most of commercial systems are based exclusively on the use of the C-band, occupying a bandwidth of roughly 5~THZ corresponding both to the minimum of the fiber loss profile, and to the amplification bandwidth of the erbium-doped fiber amplifiers (EDFA).
C+L multi-band transmission is already present in commercial systems \cite{cantono2020opportunities}, both including Raman amplification and recently  commercially developed EDFAs extended to the L-band, so enabling the exploitation of an additional 5~THz transmission bandwidth.  
Using other rare-earths than Erbium, prototype amplifier implementations have been proposed for the amplification of other bands, potentially assisted by Raman amplification, enabling the full exploitation of the entire U-to-O overall available transmission bandwidth~\cite{rapp2021optical}.

Planning and control of multi-band network require the extension of fiber transmission model for WDM lightpaths enabling full optimization and possible infrastructure sharing by software defined control based on the open physical layer abstraction \cite{icton2020curri}.
The modeling extension needs to include the variation with frequency of the fiber parameters, i.e., fiber loss, chromatic dispersion and the effective area that modifies the strengths of fiber nonlinear effects \cite{damicoJLT2022}: the Kerr effect, which generates the nonlinear interference (NLI) noise, and the stimulated Raman scattering (SRS), which induces a power transfer from higher to lower frequencies.
In particular, the SRS is the principal effect introducing multi-band interactions as the inter-band SRS-induced power transfer has a higher impact on the transmission performance than the Kerr effect \cite{9492485}.
Therefore, the SRS effect must be accurately evaluated in order to properly optimize the working point of the amplifiers in the optical line system (OLS)  in each utilized transmission band \cite{Correia:21}.
An accurate evaluation of the SRS effect on the WDM channel comb is required with respect to both the frequency and the propagation axis, $z$ as
the SRS-induced modification of the fiber loss/gain profile vs. $z$ for each frequency, $f$, with respect to the intrinsic fiber loss profile $\exp[-\alpha(f)z]$, significantly affects the amount of NLI noise.
It is lower in case of SRS-\emph{depleted} channels -- higher frequencies -- and stronger in SRS-\emph{pumped} channels -- lower frequencies \cite{cantono2018interplay,Semrau:18,9893169}.
Finally, it has been experimentally demonstrated that the approximation of transparent fiber propagation impairment on dual polarization coherent optical technologies as additive Gaussian noise is accurate also for low dispersion values \cite{6671933,ferrari2020gnpy}.
Therefore, the perturbative models evaluating the accumulated NLI noise can be extended to the entire U-to-E band and to a portion of the O-band \cite{ferrari2020assessment}.

Limiting the analysis to the C-band, the SRS effect can be accurately modeled as a spectral tilt \cite{59195}, nevertheless, this approximation is less accurate extending the transmission bandwidth, and becomes totally inaccurate when the spectral occupation exceeds the SRS efficiency peak, roughly at 13~THz \cite{9893169}.
Thus, in general, the set of ordinary differential equations (ODEs), which are the accurate mathematical model of the SRS~\cite{AGRAWAL2013,bromage2004raman}, must be solved numerically~\cite{tariq1993}.
This solution requires a non-negligible computational overhead that can be an issue due to the computation required by the transmission model for the quality of transmission estimation within the optical controller.

In this work, a perturbative solution of the SRS ODEs is proposed and validated in the optimized C+L+S and U-to-E-band transmission scenarios, where the perturbative expansions up to the 2-nd and 3-rd order, respectively, provide a high level of accuracy for the evaluation of the overall loss/gain profile along $z$ and $f$, with a maximum absolute error lower than 0.1~dB.
In conclusion, the numerical and perturbative solutions are compared in terms of accuracy and computational time with a variable launch power and an increasing total transmission bandwidth. 

The article is divided into the following sections.
In Sec.~2, the physical layer parameters involved in single mode optical fiber propagation in a multi-band context are described.
In Sec.~3, the perturbative solution inherent to the inter-channel SRS problem is reported, also introducing metrics to evaluate its accuracy.
In Sec.~4, the entire simulation system is described, including: the architecture of the network elements that make up an optical line system, the choices made in terms of launch power optimization and the different scenarios considered according to the parameters of the  physical layer.
In Sec.~5, the results obtained in terms of accuracy and computational time of the proposed perturbation solution with respect to the reference numerical method are reported and commented for each simulation.
In Sec.~6, the conclusions of this work are summarized.

\section{Physical Layer Parameters}
\label{sec:parameters}

For the sake of completeness and ease of reading, the following section is dedicated to the presentation of the physical layer parameters involved in the SRS in single-mode optical fibers, in particular highlighting their dependency with respect to the frequency.
The latter aspect is functional to have a model that accurately represents the phenomenon during optical propagation in a generic wideband transmission scenario.
In order to provide a reference for the fiber parameters descriptions, Tab.~\ref{tab:bounds} reports the frequency bounds of the each band composing the entire multi-band scenario considered in this work.
A complete description of each physical layer parameter is provided in~\cite{AGRAWAL2013} and a summary focused on a wideband transmission scenario is given in~\cite{damicoJLT2022}.

\begin{table}[]
\centering
\caption{Definition of the frequency bounds of the each band composing the considered wideband scenario, 75~GHz fixed grid.}
\label{tab:bounds}
\begin{tabular}{|c|c|c|c|}
\hline
\textbf{BAND} & \textbf{Lowest Central Frequency [THz]} & \textbf{Highest Central Frequency [THz]} & \textbf{Bandwidth [THz]} \\ \hline
\textbf{U} & \textbf{180.710} & \textbf{185.510} & \textbf{4.800} \\ \hline
\textbf{L} & \textbf{186.010} & \textbf{190.810} & \textbf{4.800} \\ \hline
\textbf{C} & \textbf{191.310} & \textbf{196.110} & \textbf{4.800} \\ \hline
\textbf{S} & \textbf{196.610} & \textbf{206.210} & \textbf{9.600} \\ \hline
\textbf{E} & \textbf{206.810} & \textbf{221.210} & \textbf{14.400} \\ \hline
\end{tabular}
\end{table}

\subsection{Loss Coefficient Function}

The power loss impairing the optical signal propagation through a fiber is taken into account by the fiber loss coefficient, $\alpha$.
The propagating signal wavelength determines the fiber attenuation~\cite{li2012optical}, and depends on the fiber composition and manufacturing process,
From a phenomenological perspective, the contributions in the wavelength range between 1.2 and 1.7~$\mu$m are the Rayleigh scattering, the violet and infrared absorption, the maxima of the OH-ion absorption at around 1.25 and 1.39~$\mu$m, and the absorption caused by phosphorous in the fiber core.
\begin{figure*}[!b]
\centerline{\includegraphics[width=0.6\linewidth]{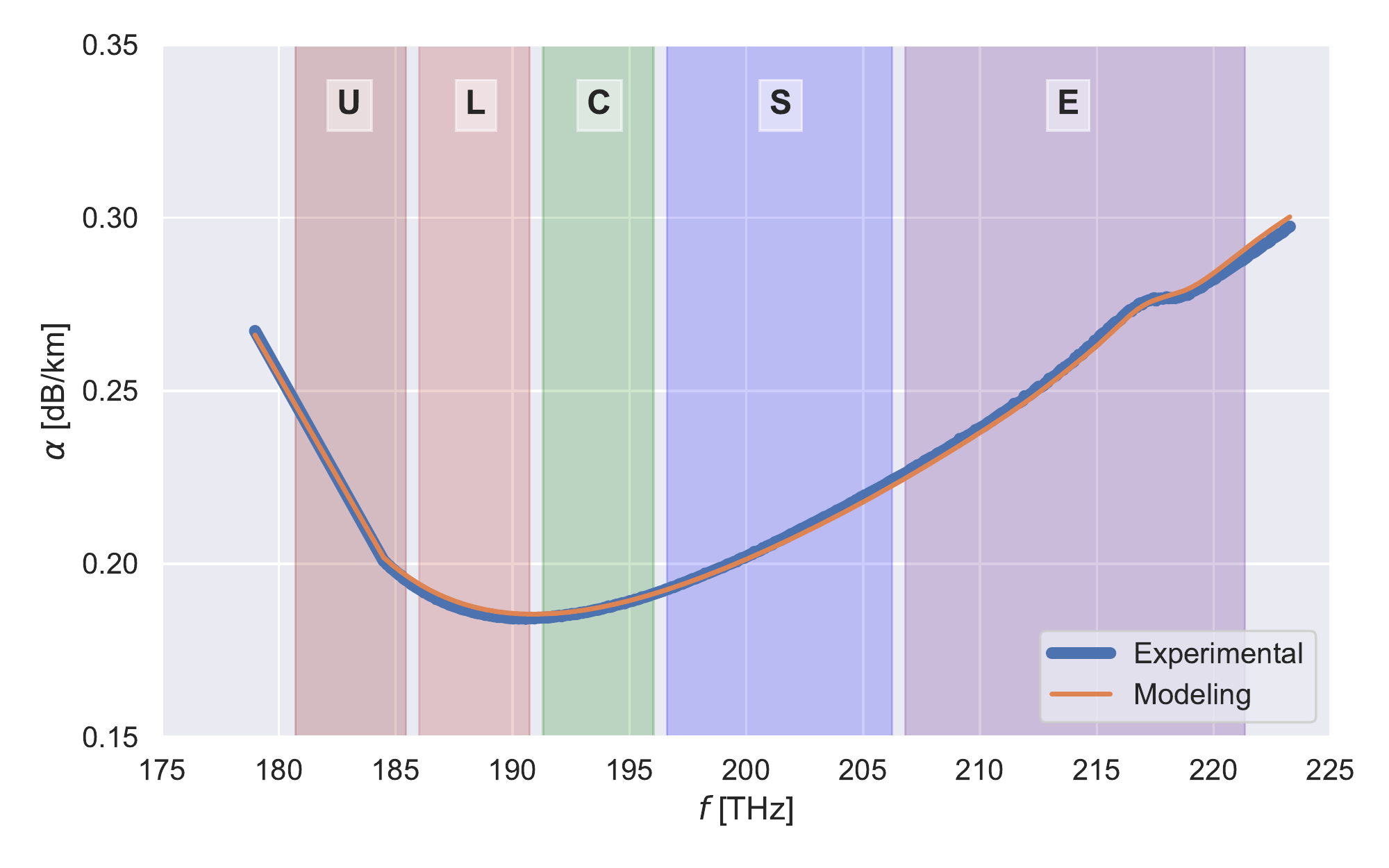}}
\caption{SSMF wideband loss coefficient profile, $\alpha(f)$.}
\label{fig:alpha}
\end{figure*}
\cite{walker1986rapid} proposed a parametric model of the loss coefficient function with regard to each phenomenological component.
The loss coefficient profile may be written as follows with regard to the optical signal wavelength, $\lambda$, and all terms written in logarithmic units (dB/km):
\begin{equation}\label{eq:fiberloss}
    \alpha(\lambda) \simeq \alpha_{\mathrm{S}}(\lambda)+\alpha_{\mathrm{UV}}(\lambda)+\alpha_{\mathrm{IR}}(\lambda)\nonumber+\alpha_{13}(\lambda)+\alpha_{12}(\lambda)+\alpha_{\mathrm{POH}}(\lambda)\:,
\end{equation}
where:
\begin{eqnarray}
    \alpha_{\mathrm{S}}(\lambda)&=& A\lambda^{-4}+B\:,\nonumber\\
    \alpha_{\mathrm{UV}}(\lambda)&=& K_{\mathrm{UV}}e^{C_{\mathrm{UV}}/\lambda}\:,\nonumber\\
    \alpha_{\mathrm{IR}}(\lambda)&=& K_{\mathrm{IR}}e^{-C_{\mathrm{IR}}/\lambda}\:,\nonumber\\
    \alpha_{13}(\lambda)&=&A_1\left(\frac{A_\mathrm{a}}{A_1} e^{\frac{-(\lambda-\lambda_\mathrm{a})^2}{2\sigma_\mathrm{a}^2}}+\frac{1}{A_1}\sum_{i=1}^{3} A_ie^{\frac{-(\lambda-\lambda_i)^2}{2\sigma_i^2}}\:\right),\nonumber\\
    \alpha_{12}(\lambda)&=&A_1\left(\frac{1}{A_1}\sum_{i=4}^{5} A_ie^{\frac{-(\lambda-\lambda_i)^2}{2\sigma_i^2}}\:\right)\:,\nonumber\\
    \alpha_{\mathrm{POH}}(\lambda)&=&A_{\mathrm{POH}}e^{\frac{-(\lambda-\lambda_{\mathrm{POH}})^2}{2\sigma_{\mathrm{POH}}^2}}\:,\nonumber
\end{eqnarray}
in turn, stand for the contributions from the Rayleigh scattering, ultraviolet, infrared, OH$-$ peak absorption, and (P)OH.
By taking into account the important elements in the C, L, and S bands, the overall model may be made simpler while ignoring the contributions from the OH-ion absorption peak at 1.25~$\mu m$ and phosphorus.
Additionally, within the interest band, the UV absorption exhibits consistent broadband behavior.
With these presumptions, it is possible to define 5 parameters:  $A$, $B$, $K_{\mathrm{IR}}$, $A_1$ and $K_{\mathrm{UV}}$ which take into account the effects of each phenomenological contribution.
In this work, a loss coefficient function retrieved from experimental measurements upon a standard single-mode fiber~(SSMF) has been used (see Fig.~\ref{fig:alpha}).
Leaving unchanged the other parameters with respect to~\cite{walker1986rapid}, a fitting procedure can be applied to the measured loss coefficient function, obtaining the following set of parameters $A=0.9192$~dB\,$\cdot\mu \mathrm{m}^{4}$\,/\,km, $B=0.0147$~dB\,/\,km, $K_{\mathrm{IR}}=5.0\cdot 10^{11}$~dB\,/\,km, $A_1=0.0043\cdot 10^{-3}$, $K_{\mathrm{UV}}=1.4655\cdot 10^{-16}$~dB\,/\,km.

\subsection{Effective Area}

The effective area may be calculated as $A_{eff}=\pi\,w^2$, where $w$ is the mode radius, which depends on the central pulse wavelength and the fiber geometry, when the mode profile of the pulse is well approximated by a Gaussian function.
In more details, the mode radius is denoted by $w=a\,/\sqrt{\ln V}$, where $a$ represents the fiber core radius and $V$ is the normalized frequency.
In the event of a minor relative index step at the core-cladding interface, $\Delta\approx(n_1-n_c)\,/\,n_1$, this may be stated as:
\begin{equation}
    V(\lambda)=\frac{2\pi}{\lambda}\,a\, n_1\sqrt{2\Delta} \;,
\end{equation}
where $n_1$ is the refractive index of the core and $n_c$ is the refractive index of the cladding.
In this work, the manufacturing fiber parameters of common SSMF values are assumed, as $a=4.2\,\,\mu \mathrm{m}$ and $n_2=2.6\cdot10^{-20}\,\,\mathrm{m}^2\,/\,\mathrm{W}$. 
In addition, the cladding refractive index and the refractive index difference with respect to the core are fixed at 1.45 and $0.31\%$, respectively.

\subsection{Raman Gain Coefficient}

The SRS is the prominent broadband nonlinear phenomena that occurs during the transmission of a WDM channel comb~\cite{bromage2004raman}.
The propagating electromagnetic field and the fiber's dielectric medium interact to create the SRS.
Since the interaction in this scenario is exclusively caused by the various channels within the spectrum, the SRS caused by the transmission of a WDM comb is sometimes referred to as Raman cross-talk in optical fiber communications.
The Raman gain coefficient, $g_R$, which quantifies the coupling between a specific pair of channels with a frequency shift of $\Delta f=f_p-f_s$, where $p$ and $s$ are the indexes of the channel at higher (pump) and lower (Stokes wave) frequencies, respectively, is the fundamental parameter that describes the regulation of the power transfer between channels during fiber propagation.
The kind and concentration of dopants in the fiber core, the reciprocal polarization state, the mode overlap between the pump and the Stokes wave, the absolute frequency of the pump, and other characteristics of the fiber and propagating channel modes all affect this coefficient.
Utilizing a reference pump at the frequency $f_{ref}$, it is feasible to determine the Raman gain coefficient profile for a single fiber~\cite{pincemin2002raman}.
In terms of optical power, the following curve may be described:
\begin{equation}\label{eq:raman_coeff}
    g_0(\Delta f, f_{ref})=\frac{\gamma_R(\Delta f, f_{ref})}{A_{eff}^{ov}(\Delta f, f_{ref})} \;,
\end{equation}
where $\gamma_R$ is the Raman gain coefficient in terms of mode intensity~(expressed in $\mathrm{m\,/\,W}$) and $A_{eff}^{ov}(\Delta f, f_{ref})$ is the effective area considering the effective area overlap between the pump and the Stokes wave.
By averaging the effective areas at the single pump and Stokes wave frequencies and assuming a Gaussian mode intensity distribution, the effective area can be calculated~\cite{rottwitt2003scaling}.

\begin{figure*}[!t]
\centerline{\includegraphics[width=0.6\linewidth]{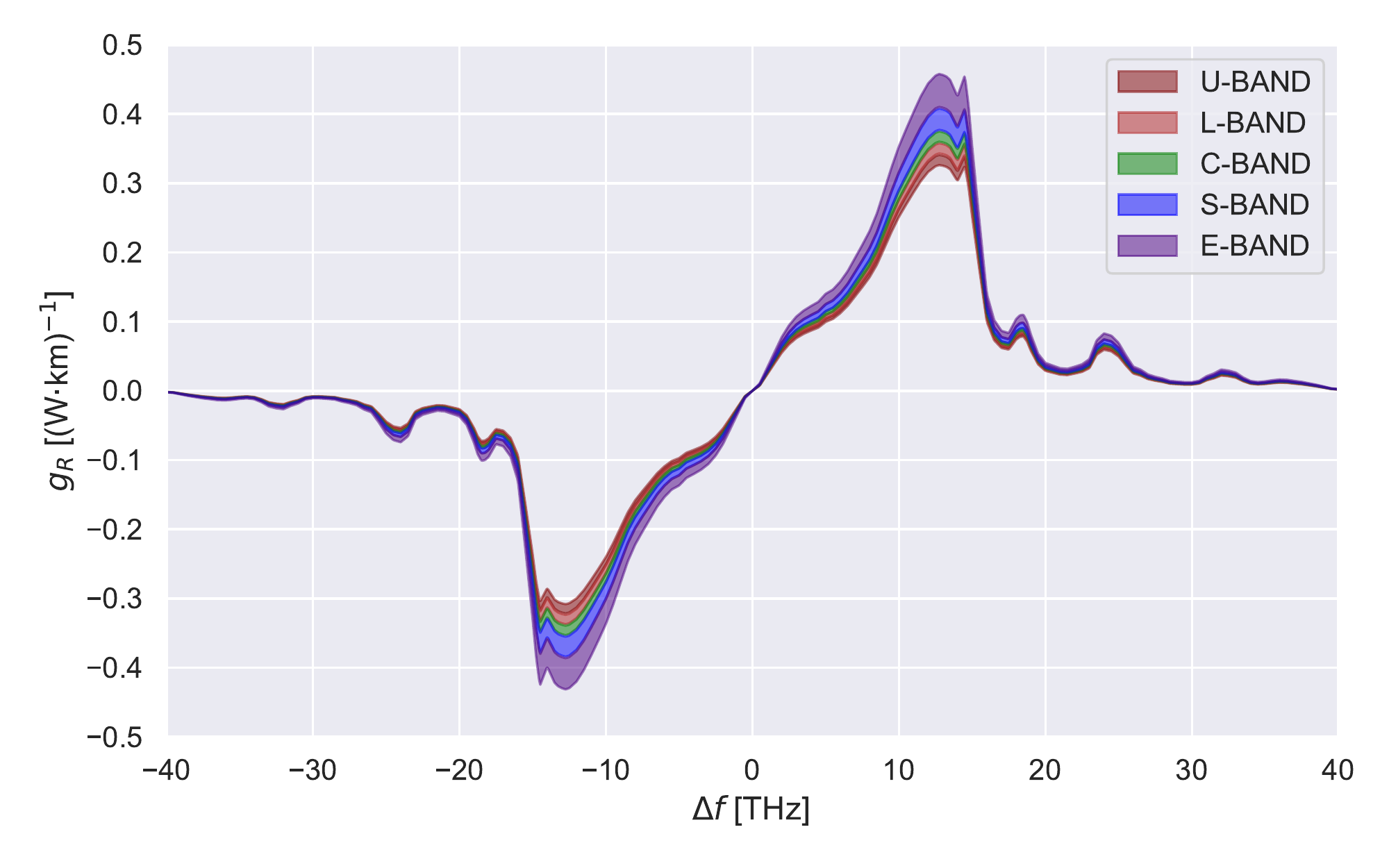}}
\caption{SSMF Raman gain coefficient profile, $g_R(\Delta f, f_{ref})$, for each frequency of the considered wideband scenario.}
\label{fig:cr}
\end{figure*}

The whole Raman gain coefficient may be modeled using the following equation in order to completely mimic optical fiber propagation and take SRS effects into account:
\begin{equation} \label{eq:cr_scaling}
    g_R(\Delta f, f_p)=k^{ps}_{pol}\,g_0(\Delta f, f_{ref})\frac{f_p}{f_{ref}}\frac{A_{eff}^{ov}(\Delta f, f_{ref})}{A_{eff}^{ov}(\Delta f, f_p)}\,\,,
\end{equation}
where the ratios between the frequencies and effective areas take into consideration the scaling of the pump and the effective area, whereas $k^{ps}_{pol}$ accounts for the reciprocal polarization state between the pump and the Stokes wave.
With regard to germanosilicate fibers, namely SSMF, the concentration of germanium in the core fiber is remarkably low, resulting in a refractive index variation of just a few hundredths of a percentage point.
In this work, the fused silica Raman gain coefficient curve reported in Fig.~\ref{fig:cr} is used, with a reference frequency of 206.185~THz.
Adding also the contribution of the vibrational (phonon) loss for higher-frequency channels with respect to the considered channel (negative half of the curve)~\cite{bromage2004raman}, the Raman efficiency profile experienced by each frequency within the wideband scenario is represented considering all the effective area scaling contributions described in the Eq.~\ref{eq:cr_scaling}.
Furthermore, as the propagating channels within the WDM comb are generally depolarized, a unitary polarization coefficient $k_{pol}$ is assumed.
In the following, for a matter of simplicity, the notation of the Raman gain coefficient is:
\begin{equation}
    g_R(\Delta f, f_{ref}) = g_R(f, f')
\end{equation}
where $f$ is the frequency of the channel under investigation and $f'$ is the interfering channel.

\section{SRS Perturbative Solution}\label{sec:perturbativeSolution}

The first order differential equation describing the Raman effect is defined on the spectral power density, $\mathcal{G}(z, f)$, as follows: 
\begin{equation}\label{eq:RamanContinuous}
    \frac{\mathrm{d}}{\mathrm{d}z} \mathcal{G}(z, f)= \left[- \alpha(f) + \int \mathrm{d}f' g_{R}(f, f') \mathcal{G}(z, f')\right] \mathcal{G}(z, f) \:.
\end{equation}
The general solution of Eq.~\ref{eq:RamanContinuous} can be decomposed as the product of the solution of the linear operator, $\mathcal{L}(z, f)$, and a nonlinear term, $\chi(z, f)$:
\begin{equation}\label{eq:RamanGeneral}
    \mathcal{G}(z, f) = \mathcal{L}(z, f) \chi(z, f)\:,
\end{equation}
given the boundary conditions:
\begin{equation}
    \left.\mathcal{G}(z, f)\right|_{z=0} = \mathcal{G}_0(f) \Rightarrow \left.\chi(z, f)\right|_{z=0} = 0\:.
\end{equation}
In particular, the solution of the linear operator is defined by the following expression:  
\begin{equation}\label{eq:RamanLinear}
    \left(\frac{\mathrm{d}}{\mathrm{d}z} + \alpha(f) \right) \mathcal{L}(z, f) = 0 \quad\Rightarrow\quad \mathcal{L}(z, f) = \mathcal{G}(z=0, f) e^{-\alpha(f) z}=\mathcal{G}_0(f) e^{-\alpha(f) z}= \mathcal{G}_0(f) \frac{\mathrm{d}}{\mathrm{d}z} \Lambda(z, f)\:,
\end{equation}
where the effective length, $\Lambda(z, f)$, is defined as the integral along $z$ of the intrinsic fiber loss:
\begin{equation}
    \Lambda(z, f) = \frac{1 - e^{-\alpha(f)z}}{\alpha(f)}
\end{equation}
On the other hand, the nonlinear term must satisfy Eq.~\ref{eq:RamanNonLinear}:
\begin{equation}\label{eq:RamanNonLinear}
    \frac{\mathrm{d}}{\mathrm{d}z} \chi(z, f) = \chi(z, f) \int \mathrm{d}f' g_R(f, f')  \mathcal{L}(z, f') \chi(z, f')\:.
\end{equation} 
A well-known exact solution of Eq.~\ref{eq:RamanNonLinear}, see \cite{Christodoulides, zirngibl1998analytical} for the details, can be derived considering a flat intrinsic loss coefficient, $\alpha(f)=\alpha \Rightarrow \Lambda(z,f)=\Lambda(z)$, and a linear Raman gain coefficient, $g_R(f,f')=- (f - f')K_R$.   
By means of these simplifications, the solution of Eq.~\ref{eq:RamanNonLinear} is:
\begin{equation}\label{RamanExact}
    \chi(z, f) = \frac{P\,e^{-f\,K_R P \Lambda(z)}}{\int \mathrm{d}f'\mathcal{G}_0(f')e^{-f'\,K_R P \Lambda(z)}},
\end{equation}
where $P=\int \mathrm{d}f \mathcal{G}_0(f)$ is the total launch power.

In general, as shown in Fig.~\ref{fig:alpha} and Fig.~\ref{fig:cr}, both the assumptions, a flat loss coefficient and a linear Raman gain coefficient, are increasingly inaccurate when the total bandwidth exceeds roughly 15~THz. 
In \cite{LasagniRaman}, a correction of Eq.~\ref{RamanExact} is proposed considering the triangular approximation of the Raman coefficient profile \cite{chraplyvy1984optical} and an interpolation of the intrinsic fiber loss coefficient. 

In this work, a perturbative approach is defined, validated and analysed.
The advantage of this approach is that, when the numerical series defined by the perturbative expansion converges, a truncated solution can be defined with an arbitrary level of accuracy, depending on the order of the truncation.
Moreover, the solution of the perturbative expansion provides a straightforward expression of the correlation between the system parameters and the final result.

First, by means of the substitution $\Gamma(z, f) = \ln\left(\chi(z, f)\right)$, Eq.~\ref{eq:RamanNonLinear} can be written as follows: 
\begin{eqnarray}\label{eq:GammaDiff}
    \frac{\mathrm{d}\Gamma(z, f)}{\mathrm{d}z} &=&  \int \mathrm{d}f' g_{R}(f, f') \mathcal{L}(z, f') e^{\Gamma(z, f')}\label{eq:GammaDiff}\\
    \Rightarrow \Gamma(z, f) &=& \int_{0}^{z}\mathrm{d}z'\int \mathrm{d}f' g_{R}(f, f')\mathcal{L}(z', f')  e^{\Gamma(z', f')}\:. \label{eq:GammaInt} 
\end{eqnarray}
In terms of the perturbative expansion, $\Gamma(z, f)$ can be formally defined as an infinite sum:
\begin{equation}\label{eq:PertExpansion}
    \Gamma(z, f) = \sum_{k=1}^{\infty} \Gamma^{(k)}(z, f)=  \Gamma^{(1)}(z, f) + \Gamma^{(2)}(z, f) + \Gamma^{(3)}(z, f) + \cdots \:,
\end{equation}
where the $k$-th order term $\Gamma^k(z, f)$ is proportional to the $k$-th power of the perturbative parameter.
By mean of this expansion, Eq.\ref{eq:GammaInt} becomes:
\begin{eqnarray}
    \Gamma(z, f) &=& \int_{0}^{z}\mathrm{d}z' \int \mathrm{d}f' g_{R}(f, f') \mathcal{L}(z', f') \prod_{k=1}^{\infty} e^{\Gamma^{(k)}(z, f)}\nonumber\\
     &=& \int_{0}^{z}\mathrm{d}z'\int \mathrm{d}f' g_{R}(f, f') \mathcal{L}_0(z', f')\prod_{k=1}^{\infty} \sum_{n=0}^{\infty} \frac{1}{n!} \left(\Gamma^{(k)}(z', f')\right)^{n}\:,
\end{eqnarray}
and the $k$-th can be expressed as follows:
\begin{eqnarray}\label{eq:RamanOrders}
    \Gamma^{(k)}(z, f) &=& \int_{0}^z \mathrm{d}z'\int \mathrm{d}f' g_{R}(f, f') \mathcal{L}(z',f') \sum_{\{n_j\}} \prod_{j=1}^{k-1} \frac{1}{n_j !}\left(\Gamma^{(k_j)}(z', f')\right)^{n_j} \:,\\
    && \forall \{n_j\}\quad \mathrm{such \; that} \quad \sum_{j=1}^{k-1} k_j\, n_j = k - 1 \: .\nonumber
\end{eqnarray}
Given Eq.~\ref{eq:RamanOrders}, successive orders can be evaluated knowing previous orders.

In particular, the first four orders are:
\begin{eqnarray}
     \Gamma^{(1)}(z, f) \!\!\!\!\!&=&\!\!\!\!\! \int_{0}^z \!\! \mathrm{d}z' \!\! \int \mathrm{d}f' g_{R}(f, f') \mathcal{L}(z', f') = \int \mathrm{d}f' g_{R}(f, f') P_{0}(f') \Lambda(z, f') \:,\label{eq:Iorder}\\
    \Gamma^{(2)}(z, f) \!\!\!\!\!&=&\!\!\!\!\! \int_{0}^z \!\! \mathrm{d}z' \!\! \int \mathrm{d}f' g_{R}(f, f') \mathcal{L}(z', f') \left[\Gamma^{(1)}(z', f')\right] \:,\label{eq:IIorder}\\
    \Gamma^{(3)}(z, f) \!\!\!\!\!&=&\!\!\!\!\! \int_{0}^z \!\! \mathrm{d}z' \!\! \int \mathrm{d}f' g_{R}(f, f') \mathcal{L}(z', f') \left[\Gamma^{(2)}(z', f') + \frac{1}{2}\left(\Gamma^{(1)}(z', f')\right)^{2}\right]\:, \label{eq:IIIorder}\\
    \Gamma^{(4)}(z, f) \!\!\!\!\!&=&\!\!\!\!\! \int_{0}^z \!\! \mathrm{d}z' \!\! \int \mathrm{d}f' g_{R}(f, f')\mathcal{L}(z', f') \left[\Gamma^{(3)}(z', f') + \Gamma^{(1)}(z', f')\Gamma^{(2)}(z', f')+ \frac{1}{3!}\left(\Gamma^{(1)}(z', f')\right)^{3}\right] \!.\label{eq:IVorder}
\end{eqnarray}

Beyond the first order, the integration in $z'$ can be analytically solved for any other orders, obtaining an expression that depends only on the system parameters and input.
As an example, the integrated solution for the second order:
\begin{eqnarray}\label{eq:analyticalSecondOrder}
    \Gamma^{(2)}(z, f) &=& \int \mathrm{d}f' g_{R}(f, f') P_{0}(f') \int \mathrm{d}f'' g_{R}(f', f'') P_{0}(f'') \nonumber\\
    &&\frac{1}{2}\left[\Lambda(z, f')\Lambda(z, f'') + \left( \frac{\alpha(f') - \alpha(f'')}{\alpha(f')\alpha(f'')} \right)\frac{1 - e^{-\left[\alpha(f') + \alpha(f'')\right]z}}{\alpha(f') + \alpha(f'')}\right]\:.
\end{eqnarray}

 It is worth to notice that, depending on the system characteristics and the specific software implementation, it may be convenient, in terms of computational cost, to perform the analytical integration in $z'$ or, instead, perform a numerical integration, maintaining an explicit expression of the previous order.
 As a matter of fact, looking at Eq.\ref{eq:analyticalSecondOrder} it can be observed that the analytical integration removes the dependency of the solution on multiple distances, $z'$, required for a numerical integration, but it implies additional integrals in the frequency space. 

In conclusion, considering the perturbative expansion Eq.~\ref{eq:PertExpansion} up to the $k$-th order, the truncated solution of Eq.~\ref{eq:RamanGeneral} is:
\begin{equation}\label{eq:truncatedSolution}
    \mathcal{G}^{(k)}(z, f)=\mathcal{L}(z,f) \exp\left[\sum_{j=1}^{k} \Gamma^{(j)}(z,f)\right]\:.
\end{equation}
Considering a total number of channels, $N_{ch}$, combined in a WDM spectrum propagating through a single fiber span, Eq.~\ref{eq:truncatedSolution} can be used to evaluate the corresponding $k$-th order power profile truncated solution: 
\begin{equation}\label{eq:powerSolution}
    \mathcal{P}^{(k)}_{ch}(z) = \int_{B_{ch}} \mathrm{d}f \:\mathcal{G}^{(k)}(z, f) = P_{ch}\:e^{-\alpha_{ch} z} \exp\left[\sum_{j=1}^{k} \Gamma_{ch}^{(j)}(z)\right]\:,
\end{equation}
where $ch\in[1,\cdots, N_{ch}]$ and $B_{ch}$ the $ch$-th channel bandwidth. 
$\alpha_{ch}$ and $\Gamma_{ch}^{(j)}(z)$ are evaluated at the channel central frequency and considered flat within $B_{ch}$, and $P_{ch} = \int_{B_{ch}} \mathrm{d}f \:\mathcal{G}_0(f)$ is the $ch$-th channel launch power.

In order to quantify the accuracy of the proposed methodology, the $k$-th order relative error can be defined in logarithmic units as follows:
\begin{equation}\label{eq:error}
    \mathcal{E}_{ch}^{(k)}(z) = 10 \log_{10}\left(\frac{\mathcal{P}_{ch} (z)}{\mathcal{P}^{(k)}_{ch}(z)}\right)= \frac{10}{\ln\left(10\right)} 
    \sum_{j=k}^{\infty} \Gamma_{ch}^{(j)}(z) \:.
\end{equation}
Eq.~\ref{eq:error} provides an explicit expression of the accuracy achieved with different orders considered in the perturbative expansion, as the estimation error is defined as the remainder left out the truncated solution of Eq.~\ref{eq:RamanContinuous}.

\section{Simulation Setup}

\subsection{Optical Line System Architecture}

\begin{figure*}[!b]
\centerline{\includegraphics[width=0.75\linewidth]{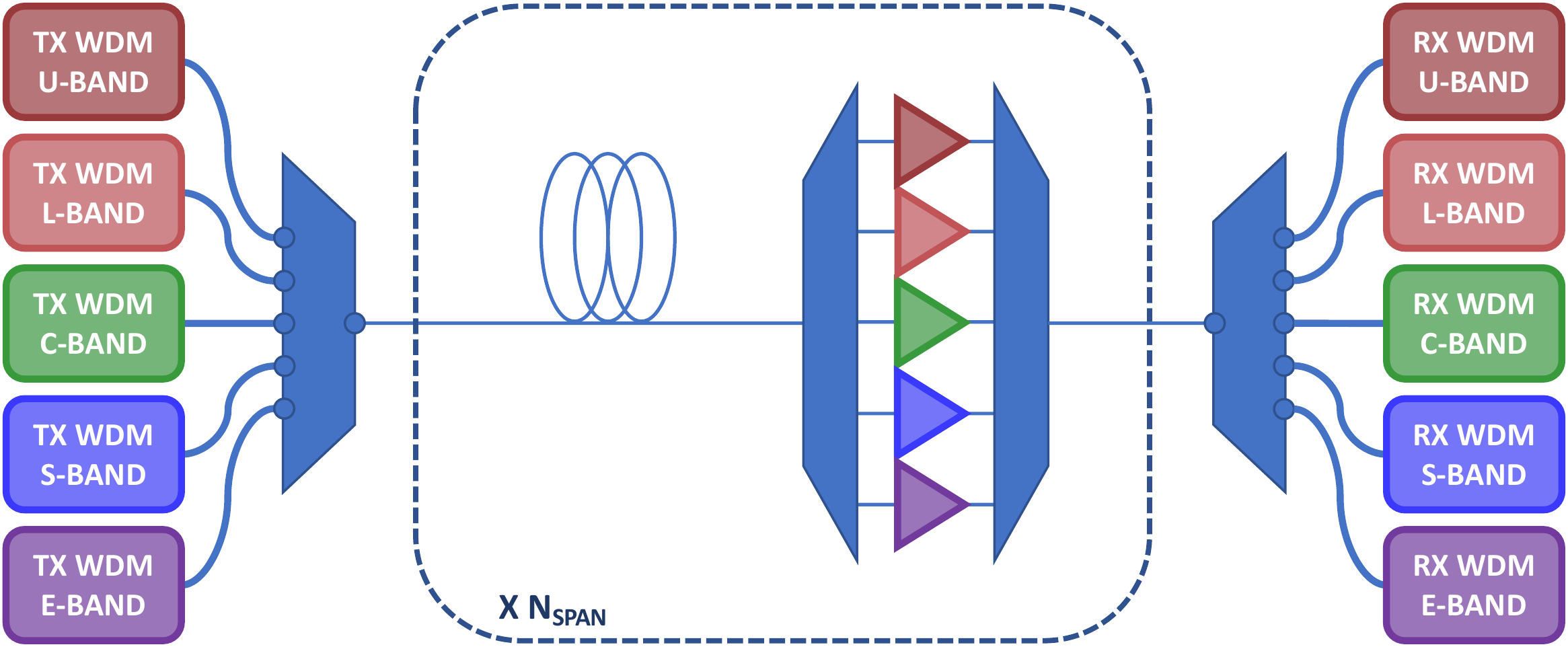}}
\caption{Sketch of the considered optical line system architecture.}
\label{fig:ols}
\end{figure*}

In order to validate and analyze the results of the proposed methodology, in this section, a realistic multi-band optical transmission system is defined and a few spectrum configuration considered. 
In particular, the truncated solution of the perturbative expansion is verified at the optimum launch power profile for each considered multi-band tranmsission scenario.

In a partially disaggregated optical network context, in which the independent routing of WDM signal is enabled by the deployment of reconfigurable optical add \& drop multiplexers (ROADMs)~\cite{curri2022gnpy}, a ROADM-to-ROADM multi-band transmission OLS is investigated, where each band is amplified by a separate and independent optical amplifiers (Fig.~\ref{fig:ols}).
The common used metric for estimating the quality of transmission (QoT) associated to a certain lightpath (LP) is the generalized signal-to-noise ratio, $\mathrm{GSNR}$.
The nonlinear signal-to-noise ratio, $\mathrm{SNR_{NL}}$, which includes the effect of the nonlinear interference noise, and the optical signal-to-noise ratio, $\mathrm{OSNR}$, which includes amplified spontaneous emission noise generated by the optical amplifiers, are the two major contributions of the GSNR.
Namely, assuming each LP as an additive and white Gaussian noise (AWGN) channel, the GSNR for a specific wavelength can be expressed as:
\begin{equation}\label{eq:ols_gsnr}
    \mathrm{GSNR} = \left(\mathrm{OSNR}^{-1} + \mathrm{SNR}^{-1}_{\mathrm{NL}}\right)^{-1} \:,
\end{equation}
where $\mathrm{OSNR}^{-1}$ and $\mathrm{SNR}^{-1}_{\mathrm{NL}}$ can be written as the sum of the separate noise contributions generated in each crossed span.

\subsection{Launch Power Optimization}\label{subsec:PowerOptimization}

The optical line controller responsible for the OLS operation defines the working points of each amplifier to optimizing the GSNR of the OLS.
Specifically, a convenient optimization strategy can be defined enforcing a maximized and uniform GSNR spectral distribution on each transmission band.
As a result, the chosen optimization criteria in this study is based on the definition of the launch power profile that simultaneously provides the highest average per-band GSNR value and is sufficiently flat on each single transmission band and, in general, throughout the entire transmitted spectrum.    

The proposed optimization approach does not need any extra hardware because the optimal launch power is achieved by adjusting the gains (or output power) and tilt settings of each optical amplifier.
In conclusion, the number of variables to be optimized is twice the number of bands, $N_B$, of the considered multi-band transmission scenario (the pair of gain/tilt values for each $n$-th amplifier along the OLS), and the objective function to be maximized is:
\begin{equation}
    \max \left(\frac{1}{N_B}\left[\,\sum_{n=1}^{N_B}\left(\overline{\mathrm{GSNR}_n} - \sigma_{\mathrm{GSNR}_n}\right)\right] - \sigma_{\{\overline{\mathrm{GSNR}_1}, ... \overline{\mathrm{GSNR}_{N_B}}\}}\right)
\end{equation}
where $\overline{\mathrm{GSNR}_n}$ is the average GSNR value of the $n$-th band, $\sigma_{\mathrm{GSNR}_n}$ is the GSNR standard deviation of the $n$-th band and $\sigma_{\{\overline{\mathrm{GSNR}_1}, ... \overline{\mathrm{GSNR}_{N_B}}\}}$ is the standard deviation computed on the set of all GSNR average values.

A stochastic heuristic optimization method based on an evolutionary approach has been leveraged in to solve this optimization problem; the covariance matrix adaptation evolution strategy (CMA-ES) is applied as optimization algorithm.

\begin{table}[!t]
\centering
\caption{Optical amplifier noise figure values used in the considered wideband scenario.}
\label{tab:noise_figure}
\begin{tabular}{|c|c|c|c|c|c|}
\hline
\textbf{} & \textbf{U} & \textbf{L} & \textbf{C} & \textbf{S} & \textbf{E} \\ \hline
\textbf{NF [dB]} & \textbf{6.0} & \textbf{6.0} & \textbf{5.5} & \textbf{7.0} & \textbf{7.0} \\ \hline
\end{tabular}
\end{table}

\subsection{Analysed Scenarios}

\begin{figure}[!t]
\centering
\begin{subfigure}{.45\textwidth}
\includegraphics[width=\linewidth]{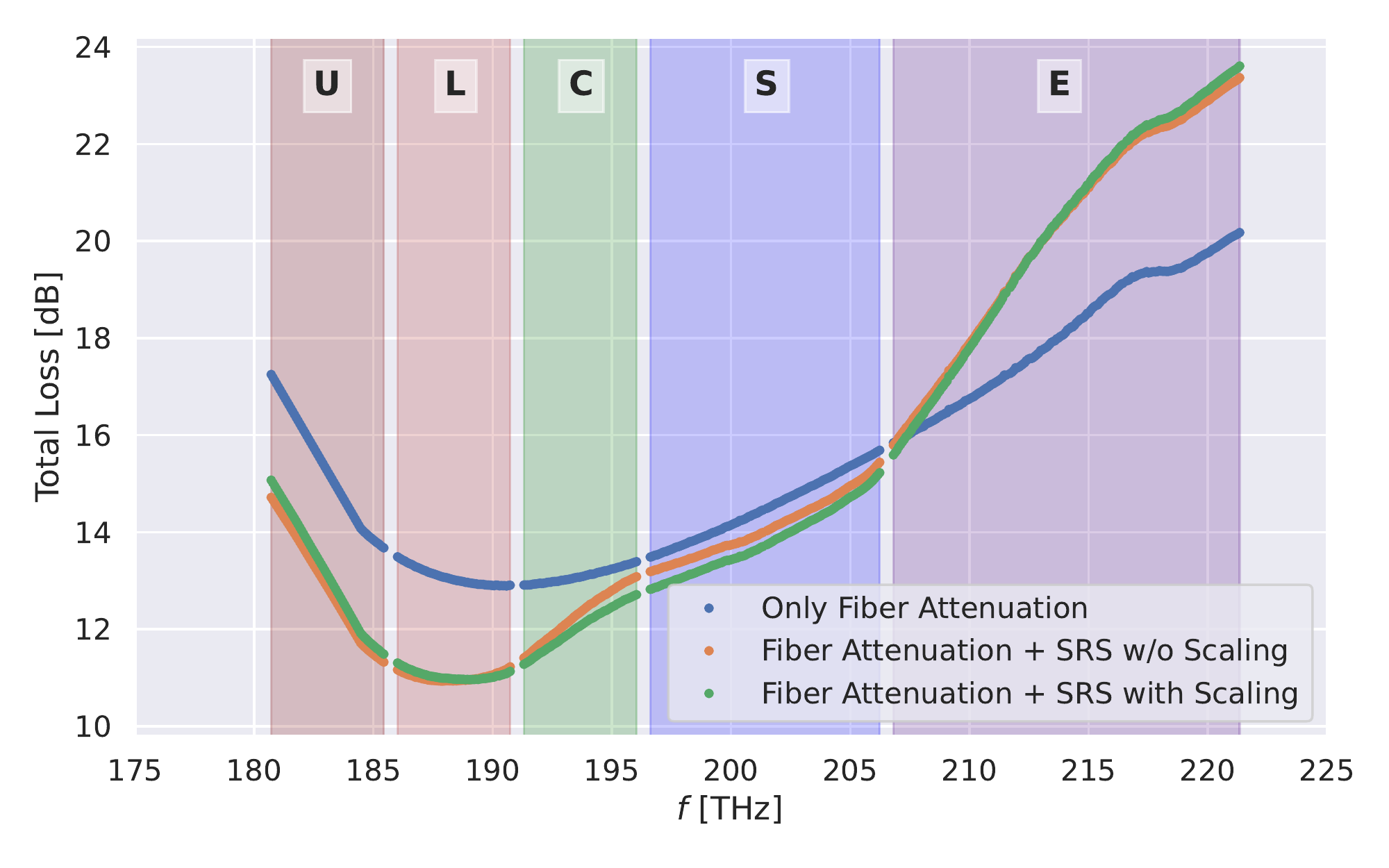}
\caption{Realistic fiber loss coefficient in a wideband transmission scenario.}
\label{fig:scaling_realistic}
\end{subfigure}%
\hfill
\begin{subfigure}{.45\textwidth}
\includegraphics[width=\linewidth]{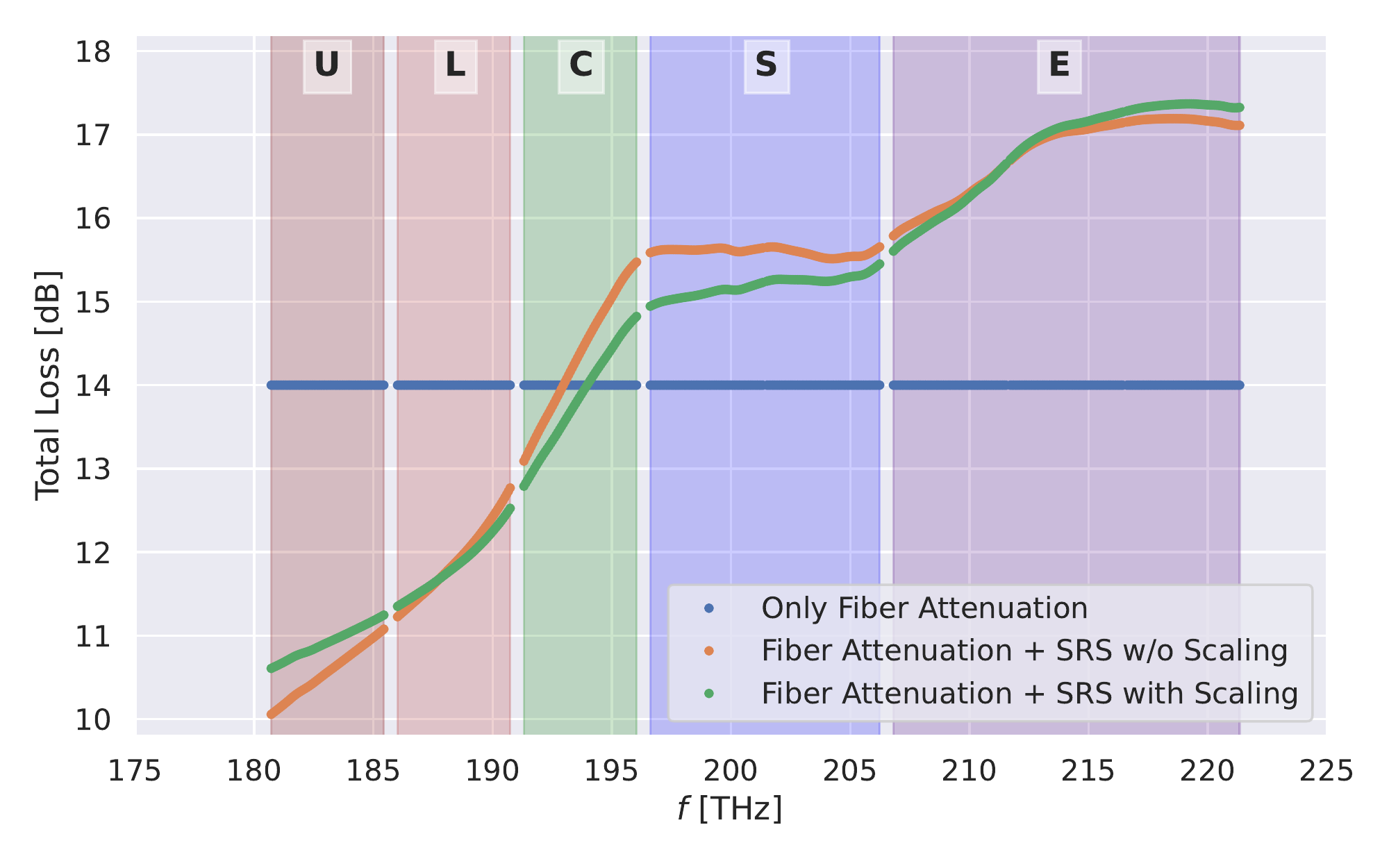}
\caption{Ideal fiber loss coefficient in a wideband transmission scenario.}
\label{fig:scaling_ideal}
\end{subfigure}
\end{figure}



For the purpose of this work, a periodic multi-band OLS of 10 spans is considered.
The assumed values of noise figure, $\mathrm{NF}$, for each optical amplifier type are reported in Tab.~\ref{tab:noise_figure} according to the corresponding band \cite{ferrari2020assessment}.
The fiber spans are 70~km long and are characterized by the realistic wideband parameter description reported in Sec.~\ref{sec:parameters}.
The transmitted signal is implemented according to the 400G standard: Each channel carries a dual polarizatio (DP) 16-QAM (quadrature amplitude modulated) signal with a symbol rate of 64\,GBaud and a slot width of 75\,GHz.

In this framework, a cutting-edge C+L+S-band transmission scenario is considered along with a more future looking U-to-E-band transmission scenario in order to perform a solid validation of the proposed methodology.
In both cases, the optimal launch power has been evaluated with the procedure described in Sec.~\ref{subsec:PowerOptimization}. 

Additionally, an ideal flat loss coefficient profile at 0.2~dB\,/\,km is considered in order to separately analyze the effect of the SRS effect on the power profile along the fiber.
In this case, a flat launch power of -1~dBm per channel is considered.
The impact of the real and ideal fiber loss coefficient profiles in terms of total attenuation along a single fiber span is shown in Figs.~\ref{fig:scaling_realistic} and~\ref{fig:scaling_ideal}.
In these figures, also the effect of the effective area scaling is shown; its contribution produces variation within each transmission band in a range of 0.5~dB, suggesting that it is necessary to consider this phenomenon mainly for the OSNR estimation especially after the propagation of a WDM comb through a considerable number of sections.
Even if this effect turns out to be of secondary importance, in a wideband context its accumulation across the various spans can generate additional inaccuracy.

\section{Results}

First, for all the transmission scenarios, the solution of Eq.~\ref{eq:RamanContinuous} has been solved using the numerical integration defined in~\cite{tariq1993}, which provides an accurate reference if the position increments are small enough; for the validation purpose, a constant step of 0.8~m has been chosen, providing an accuracy of the evaluated power profile along the fiber of 0.001~dB, for all the frequencies of the transmitted spectra.
Then, the reference is compared with a truncated solution of the perturbative expansion presented in Sec.~\ref{sec:perturbativeSolution}.
Given the optimal launch powers of all the considered transmission scenarios, the perturbative expansion Eq.~\ref{eq:PertExpansion} is convergent.
In particular, the successive orders are monotonously decreasing and, moreover, considering the $k$-th order truncated solution, the infinite sum of the remainder orders converges.
Therefore, an arbitrarily small relative error, $\mathcal{E}_{ch}^{(k)}(z)$, for all the channels of the propagated spectrum can be achieved considering the proper $k$-th order of the solution.

\begin{figure*}[h]
\centerline{\includegraphics[width=\linewidth]{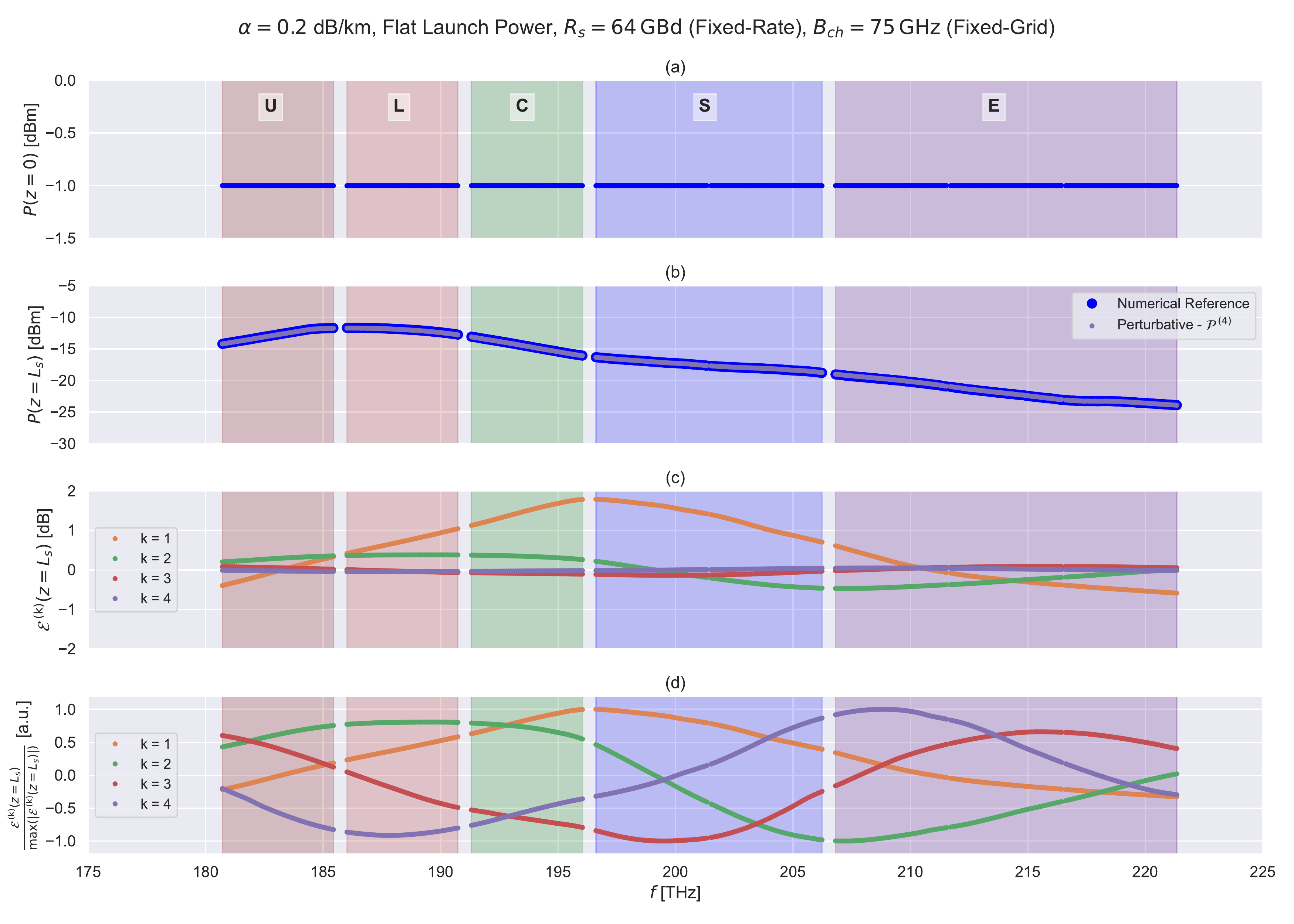}}
\caption{Ideal flat fiber loss profile simulations. In particular, (a) Flat launch power at -1\,dBm per channel. (b) Comparison of the numerical reference and the 4-th order perturbative solution. (c) Relative error up to the 4-th order. (d) Normalized relative error up to the 4-th order.}
\label{fig:orders}
\end{figure*}
In Fig.~\ref{fig:orders}-b, the $4$-th truncated solution of Eq.~\ref{eq:powerSolution} is compared with the reference evaluation in the case flat fiber loss profile and $P_{ch}$ (Fig.~\ref{fig:orders}-a), at the fiber termination, $L_s$.
In Fig.~\ref{fig:orders}-c, it is shown that increasing the order of the solution, an increasing accuracy is obtained, until the arbitrary tolerance of 0.1\,dB is achieved.
Finally, in Fig.~\ref{fig:orders}-d, the normalized error $\mathcal{E}_{ch}^{(k)}(L_{s})/\max\left(|\mathcal{E}_{ch}^{(k)}(L_{s})|\right)$ up to the 4-th order is shown.
As expected, it can be observed that the $k$-th relative error has exactly the symmetry in frequency of the $k+1$-th order, which it is its most significant term.
Therefore, odd orders have an even error function in frequency and \textit{vice versa}, demonstrating that the proposed perturbative expansion is exact at every order.

Further analysis on the formal expansion Eq.~\ref{eq:PertExpansion} are out of the scope of this study and will be investigated in future publications.
A heuristic effective and conservative estimation of the proper order required to achieved at least a given tolerance, $\tau$, has been validated for a set of increasing U-to-E transmission bandwidth and an increasing flat launch power per channel, $P_{ch}\in[-4, 2]$\,dBm, resulting in a total power at the fiber input for the full U-to-E scenario of 23.1 and 29.1~dBm, respectively; this evaluation provides the correct order or at most the successive, guaranteeing the required accuracy.
Due to the complex spectral shape of the Raman coefficient profile, distinct orders have different interactions with the power profile along the fiber, therefore, the proposed estimation procedure has to be perform after the calculation of each order and it is expressed by the following inequality:
\begin{equation}\label{eq:orderEstimation}
    \left|\mathcal{E}_{ch}^{(k)}(z)\right| \leq \frac{10}{\ln(10)} \left[\exp\left(\theta^{(k)}\right) - \sum_{j=0}^{k} \frac{\left(\theta^{(k)}\right)^{j}}{j!}\right]\leq \tau\:,\quad \mathrm{with} \quad \theta^{(k)} = \sqrt[k]{k! \: \max\left(\left|\Gamma^{k}_{ch}(z)\right|\right)}\:.
\end{equation}

\begin{figure*}[!t]
\centerline{\includegraphics[width=\linewidth]{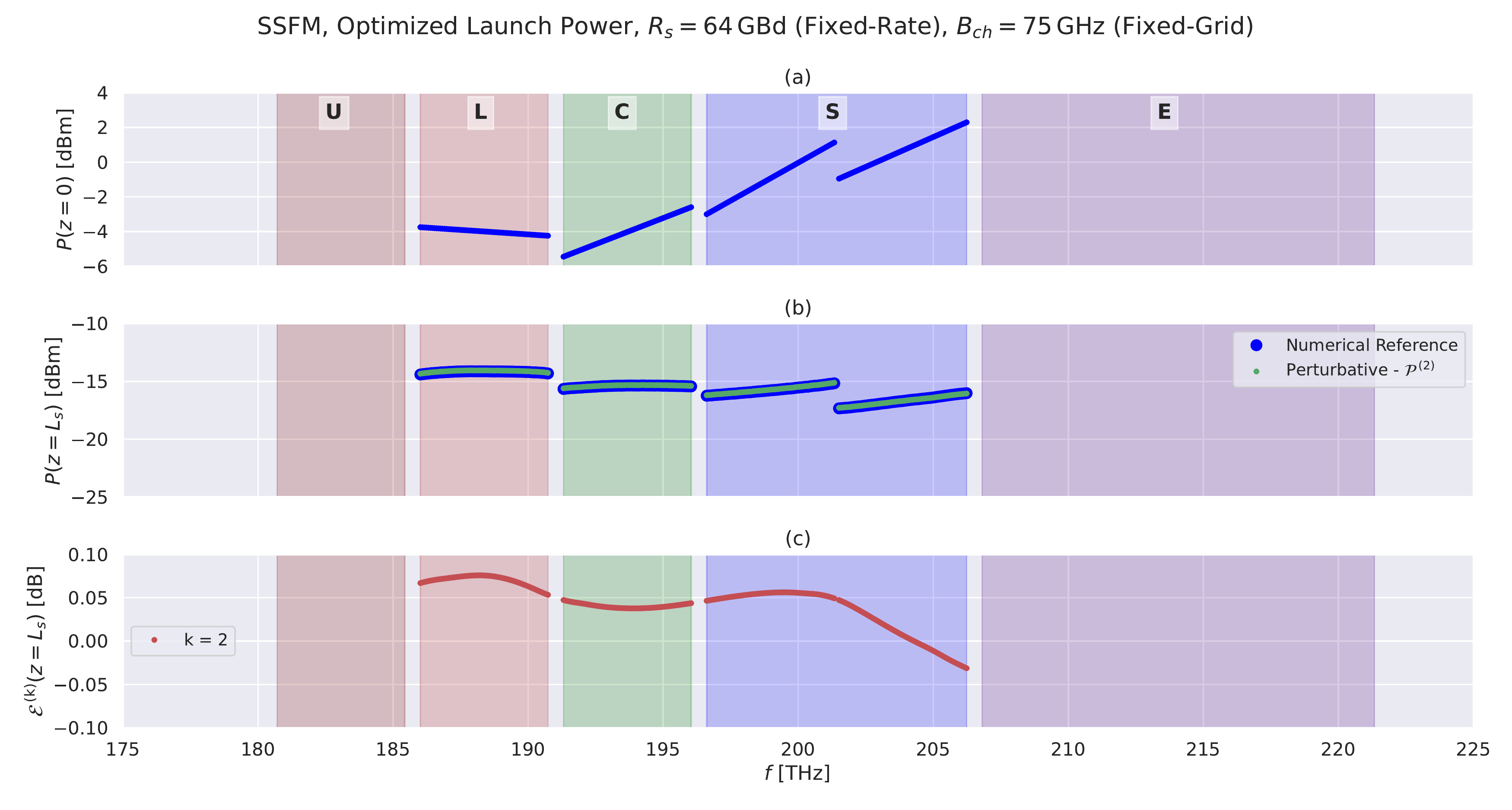}}
\caption{Realistic fiber parameters simulations, C+L+S-band transmission scenario. In particular, (a) Optimal launch power. (b) Comparison of the numerical reference and the 2-nd order perturbative solution. (c) Relative error up to the 2-nd order.}
\label{fig:optimum_cls}
\end{figure*}
\begin{figure*}[!t]
\centerline{\includegraphics[width=\linewidth]{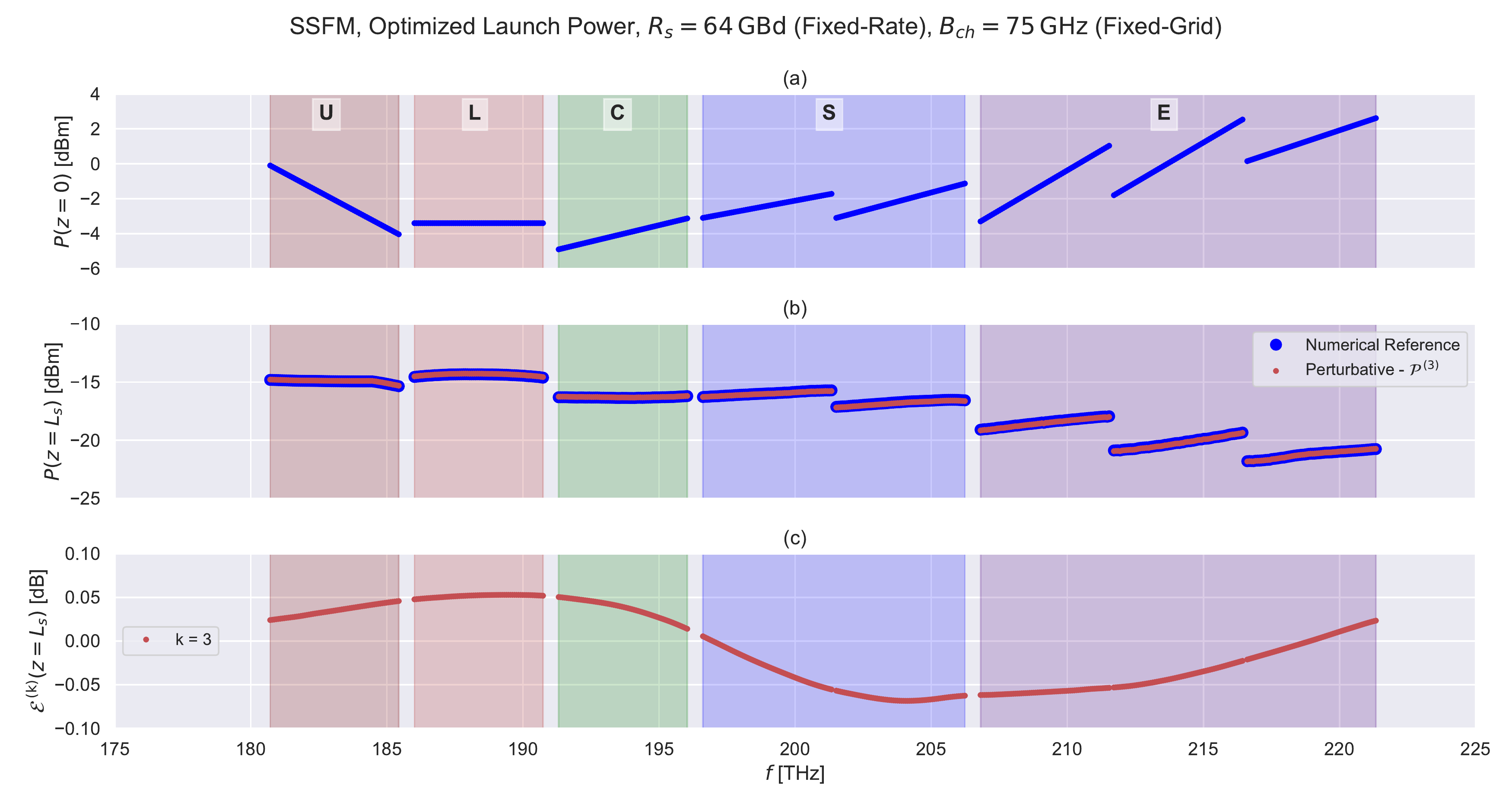}}
\caption{Realistic fiber parameters simulations, U-to-E-band transmission scenario. In particular, (a) Optimal launch power. (b) Comparison of the numerical reference and the 3-rd order perturbative solution. (c) Relative error up to the 3-rd order.}
\label{fig:optimum_all}
\end{figure*}
Using Eq.~\ref{eq:orderEstimation}, with a defined tolerance of 0.1\,dB, the truncated solution at the proper order of Eq~\ref{eq:powerSolution} has been evaluated for all the investigated realistic transmission scenarios; for both cases the required tolerance is achieved at the exact order evaluated by means of Eq.~\ref{eq:orderEstimation}.
In Fig.~\ref{fig:optimum_cls} and Fig.~\ref{fig:optimum_all}, the launch power profile, the reference and evaluated power profile at the fiber termination and the relative error are reported for the C+L+S-band and U-to-E transmission scenarios, respectively.

Finally, the proposed methodology enables a faster implementation of the SRS solver for all the transmission scenarios with respect to the numerical integration method, given a certain tolerance value.
As an example, the integral solution of each order, as Eq.~\ref{eq:Iorder}-~\ref{eq:IVorder}, can be integrated numerically in order to obtain the perturbative solution without evaluating an explicit form, as Eq.~\ref{eq:analyticalSecondOrder}, for each perturbative order.
Remarkably, the position increment required to achieved a given accuracy spatially integrating the perturbative orders, roughly tens of km if $\tau=0.1$~dB, is significantly larger then the position increment required by the numerical solution to achieve the very same accuracy, roughly 0.1-1\,km.
Therefore, the perturbative solution required a significantly lower computational effort to achieve the same result of the numerical solution.
Regarding the explicit expression of each order, as anticipated in Sec.~\ref{sec:perturbativeSolution}, it is not always convenient in terms of computational cost.
This is due to the high number of channels involved in a wideband scenario and can be overcome considering a lower number of equivalent macro channels in place of the real propagated channels, assuming that the variations of the intrinsic fiber loss, the Raman coefficient profile and the power spectral density are negligible within the macro channel bandwidths.
Further consideration on this aspects will be addressed in a future publication.

\begin{figure*}[!t]
\centerline{\includegraphics[width=0.6\linewidth]{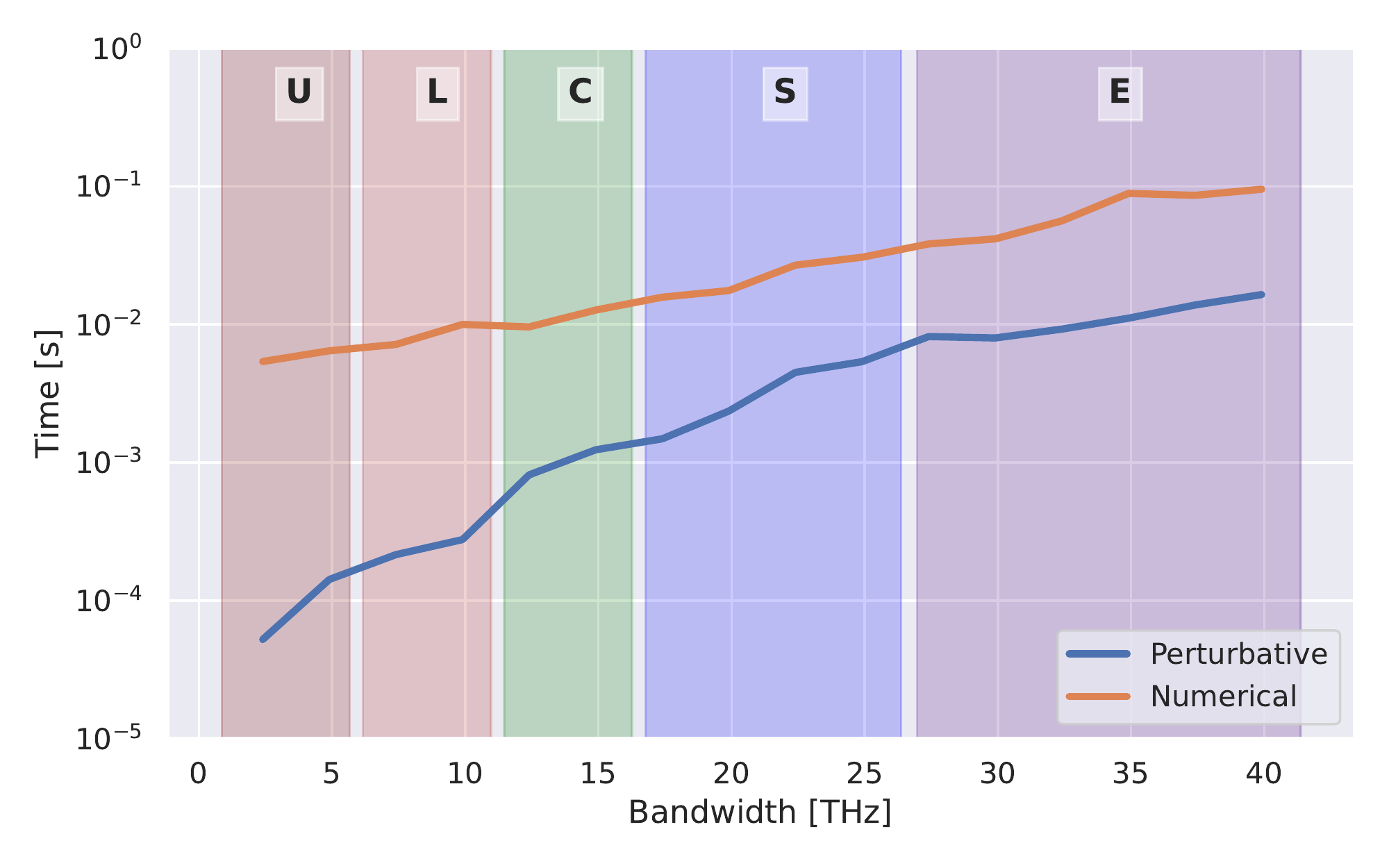}}
\caption{Computational time of the perturbative and numerical solution, respectively, evaluated for an increasing total bandwidth of the transmitted spectrum with a flat power of -1\,dBm per channel.}
\label{fig:time_mae}
\end{figure*}

In this work, an increasing total transmission bandwidth is considered from 2.5 to 40\,THz starting from the first portion of the U-band to the last portion of the E-band, with a step of 2.5\,THz. 
For all this spectra, a fixed power per channel of -1\,dBm has been set and the proper order and position increment for the spacial integration has been evaluated for both the perturbative and numerical solution in order to obtain a fixed 0.1\,dB tolerance.
The resulting computational times are reported in Fig.~\ref{fig:time_mae}, where it can be observed that the perturbative solution perform at least one order of magnitude better than the numerical solution.

\section{Conclusion}

In this study, a perturbative expansion describing the inter-channel SRS is presented and analysed.
The proposed methodology enables the estimation of an arbitrary accurate solution when the proper perturbative order is evaluated, and it provides a direct insight of the relation between the SRS effect and the fiber or spectrum parameters.
The proposed perturbative solution is validated in wideband transmission scenarios including the C+L+S-band and U-to-E-band transmission scenario.
Additionally, a heuristic effective and conservative procedure for the estimation of the proper perturbative order required to achieve a given tolerance is provided and discussed.
Finally, the benefit in terms of computational time of the proposed methodology is demonstrated on increasing total bandwidth transmission scenarios.

In future publications, the convergence of the perturbative expansion will be investigated to further extent, along with an optimal software implementation of the proposed methodology.
Also, the explicit expressions of the first order solutions of Eq.~\ref{eq:RamanContinuous} will be considered for an effective definition of the generalized Gaussian noise model estimating the nonlinear interference noise in wideband transmission scenarios.

\printbibliography

\end{document}